\documentclass[10pt,twocolumn,letterpaper]{article}

\usepackage{cvpr}              
\usepackage{lineno}
\usepackage{float}
\usepackage[normalem]{ulem} 
\usepackage{amsmath}
\usepackage{bbm}
\usepackage{graphicx}
\usepackage{multirow}
\usepackage{siunitx}
\usepackage{overpic}
\usepackage{algorithm, algorithmic}
\newcommand{\E}{\mathbb{E}}

\def\cI{{\mathcal I}}

\newcommand{\normShort}[1]{\Vert#1\Vert}

\newlength{\paragraphSpace}
\newcommand{\Paragraph}[1]{\vspace{\paragraphSpace}\noindent \textbf{#1}}

\definecolor{cvprblue}{rgb}{0.21,0.49,0.74}
\usepackage[pagebackref,breaklinks,colorlinks,allcolors=cvprblue]{hyperref}

\def\paperID{*****} 
\def\confName{CVPR}
\def\confYear{2025}
\def\LambdaNaive {\widehat{\Lambda}_{ij, \rm naive}}
\def\LambdaSingle{\widehat{\Lambda}_{\rm single}}
\def\LambdaMulti {\widehat{\Lambda}_{\rm multi}}
\def\PNaive {\widehat{q}_{ij, \rm naive}}
\def\PSingle{\widehat{q}_{ij, \rm single}}
\def\PMulti {\widehat{q}_{ij, \rm multi}}

\title{Image Reconstruction from Readout-Multiplexed Single-Photon Detector Arrays}

\author{Shashwath Bharadwaj, Ruangrawee Kitichotkul, Akshay Agarwal and Vivek K Goyal\\
Boston University\\
{\tt\small shash@bu.edu}
}

\begin{document}
\maketitle

\begin{abstract}
Readout multiplexing is a promising solution to overcome hardware limitations and data bottlenecks in imaging with single-photon detectors. Conventional multiplexed readout processing creates an upper bound on photon counts at a very fine time scale, where frames
with multiple detected photons must either be discarded or allowed to introduce significant bias. We formulate multiphoton coincidence resolution as an inverse imaging problem and introduce a solution framework to probabilistically resolve the spatial locations of photon incidences. Specifically, we develop a theoretical abstraction of row--column multiplexing and a model of photon events that make readouts ambiguous. Using this, we propose a novel estimator that spatially resolves up to four coincident photons.
Monte Carlo simulations show that our proposed method increases the peak signal-to-noise ratio (PSNR) of reconstruction by 3 to 4 dB compared to conventional methods under optimal incident flux conditions.
Additionally, this method reduces the required number of readout frames to achieve the same mean-squared error as other methods by a factor of $\sim 4$.
Finally, our method 
matches the Cram{\'e}r--Rao bound for detection probability estimation for a wider range of incident flux values compared to conventional methods. 
While demonstrated for a specific detector type and readout architecture, this method can be extended to more general multiplexing with different detector models.

\end{abstract}    
\section{Introduction}
\label{sec:intro}

Single-photon detectors have been widely studied for use in applications like biological imaging~\cite{bruschini2019single,wang2022emerging}, lidar~\cite{rapp2020advances,guan2022lidar}, and quantum optics~\cite{ceccarelli2021recent}
due to properties such as single-photon sensitivity and high-precision time-tagging of photon arrivals. Single-photon avalanche diodes (SPADs) have been of particular interest due to compatability with CMOS manufacturing techniques and room temperature operation~\cite{cusini2022historical}. However, they have low detection efficiencies (especially at mid infrared wavelengths), which means that only a small fraction of incident light is detected as photons. Superconducting nanowire single-photon detectors (SNSPDs) are emerging devices that offer near-unity quantum efficiency across a range of wavelengths, low dark counts, and low jitter, making them compelling alternatives to SPADs~\cite{you2020superconducting}.

However, the requirement of cryogenic cooling to maintain superconductivity has largely restricted their use to laboratory experiments. 

Single-photon imaging in typical commercial applications requires \emph{arrays} of single-photon detectors~\cite{piron2020review}. While megapixel SPAD arrays have already been deployed in mass consumer products~\cite{morimoto2021megapixel,bonannophysical}, high data transfer rates from these arrays for further processing remain a bottleneck. For instance, typical room-scale ranging experiments with SPAD array-based single-photon lidar can have data generation rates of the order of gigabits per second when the array resolution approaches the megapixel scale. 

It is attractive to compress the readout to less than a stream of all digitized detection times. Methods like in-pixel histogramming \cite{Hutchings2019}, histogram compression through sketching \cite{Gribonval2021}, equidepth histogramming \cite{10210115,sadekar2024single}, and differential readouts \cite{White2022ISCAS} have been proposed to address high data transfer rates. These approaches to avoid the cost of high temporal resolution are complementary to our emphasis on transverse spatial dimensions in this work.

Implementing high spatial resolution SNSPD arrays is a significant challenge. As the number of pixels in the array grows, the heat load introduced by reading out each pixel individually becomes incompatible with the cryogenics of the system~\cite{esmaeil2021superconducting}. To mitigate this, recent efforts have focused on addressing groups of pixels in the readout, starting with  Allman \textit{et al.}~\cite{allman2015near} who introduced row--column readout multiplexing. In this mechanism, readouts indicate photon detection at each row and each column of the array instead of every pixel, which reduces the required number of readout lines for an $n \times n$ array from \(n^2\) to \(2n\). Wollman \textit{et al.}~\cite{wollman2019kilopixel} extended this scheme to demonstrate the first kilopixel SNSPD array. McCaughan \textit{et al.}~\cite{mccaughan2022thermally} introduced the thermally coupled imager with time-of-flight multiplexing, where separate readout lines for each row and column are replaced with a single bus for all rows and all columns. This makes the required number of readouts independent of the array size. Recently, Oripov \textit{et al.}~\cite{oripov_superconducting_2023} leveraged this method to demonstrate a 400,000 pixel superconducting camera, showing the viability of this scheme to approach the sizes of commercial imaging arrays.

Despite the gain in scaling array sizes, readout multiplexing requires the incident photon flux to be very low for unambiguous signal reconstruction, limiting the array's photon count rate. When two or more photons are incident on the array within the period of a single acquisition, row--column readouts can give an ambiguous set of candidate locations for their incidences. Traditional reconstruction techniques~\cite{wollman2019kilopixel} do not distinguish between these locations and assume equal probabilities of detection at each candidate pixel. This results in the misattribution of photon counts to locations where no photon incidences occurred. Alternatively, since readouts with a single detected photon provide the exact row and column indices where the photon was incident on the array, ambiguous readouts with multiple detections can be discarded for signal reconstruction. However, this leads to high variance in the reconstruction (especially at high incident fluxes) and underutilization of the spatial dimensions of the array by requiring most readouts to contain detections at a single pixel. Resolving the spatial locations of multiphoton coincidences using multiplexed readouts is an ill-posed inverse problem.

Here, we propose a solution framework to leverage most of the measured data to reduce the mean-squared error of image reconstruction. Specifically, we develop a \textit{multiphoton estimator} that redistributes photon counts from ambiguous readouts to candidate pixels such that an approximate likelihood of observations is maximized. A key choice for our algorithm design is the omission of spatial priors in order to not obscure the source of performance gain. Spatial priors can be used in conjunction with the developed method for further improvements.

Monte Carlo simulations show that our proposed method increases the peak signal-to-noise ratio (PSNR) of reconstruction by 3 to 4 dB compared to conventional methods under optimal incident flux conditions.

Further, the optimal incident photon flux for our method is $\sim$1.4 photons per readout which is 0.5 to 1.1 photons per readout higher than conventional methods. We also demonstrate that our estimator matches the Cram{\'e}r--Rao Bound (CRB) for the estimation of photon detection probabilities at each pixel of an array across a range of incident photon flux values. Finally, our method achieves a factor of $\sim$4 reduction in the required number of readout frames for the same mean-squared error of image reconstruction. These results demonstrate that our proposed estimator is well-suited for high-flux, low-latency imaging.

\section{Related Work}
\label{sec:rel_work}
\paragraph{Image reconstruction from multiplexed readouts.}
Compressed sensing methods have been explored for the design of multiplexed readout architectures in areas like positron emission tomography (PET) \cite{6471237,olcott2011compressed, goertzen2013design, lee2017development}, biological imaging \cite{brennan2022multiplexed,rasmussen20133} and gamma-ray imaging \cite{bonifacio2012modeling,boo2021row}. While readout multiplexing for specific imaging applications has been extensively studied \cite{jensen2022anatomic, callens2021analysis}, the focus of these works is on the modeling of optical and electrical properties of the imaging system such as the shape of the wavefront of the detected light \cite{rasmussen20153, ben2015systematic}, coded apertures for rejecting multipixel events \cite{boo2021row} and IC design for resistor multiplexing \cite{goertzen2013design}. The explicit study of individual photon events measured by single-photon detectors and their impact on the multiplexed readouts does not yet exist. Our work introduces a theoretical study of the readouts based on the combinatorics of photon events and an approximate maximum likelihood estimation of photon detection probabilities at each pixel. We also analyze the performance of a novel multiphoton estimator with variations in imaging parameters. Our work is closest to the analysis presented by van den Berg \emph{et al.} \cite{van2013single}, where group-testing inspired surface codes are studied for deterministic multiphoton coincidence resolution in time-to-digital converter readouts from SPAD arrays. However, our work specifically addresses intensity imaging with SNSPD arrays where a small probability of error is tolerable. We provide a probabilistic model of the readouts and solutions for image reconstruction from a specific device architecture for up to four coincident photons.

\Paragraph{Imaging with SNSPD arrays.}
While large-scale SNSPD array development is still in the nascent stages, several works have conducted proof-of-principle imaging experiments with prototypes. Zhao \emph{et al.}\ introduced a delay-line-based superconducting imager with an effective pixel count of $\sim$590~\cite{zhao2017single}.
This method inspired the thermally coupled imager~\cite{mccaughan2022thermally}, which in turn resulted in the 400,000-pixel camera by Oripov \emph{et al.}~\cite{oripov_superconducting_2023}. Subsequent efforts to use SNSPD arrays for imaging \cite{guan2022snspd, wang2023image, kong2023readout} involve intricate hardware control such as bias-current sweeping, use of idle pixels, and geometric design of the nanowire structure. Our solution is purely computational and does not require additional hardware modifications. In addition, the modeling of incident photon events and their multiplexed readouts can be extended to several detector models in high-energy physics or particle detection. 

While the method described in this work provides a framework to resolve an arbitrarily large number of coincident photons, we limit the discussion in the next sections to up to four coincident photons to make the demonstrated solutions computationally tractable and easier to illustrate and analyze. 

\begin{figure*}
\hspace{-1.5cm}
{
\hspace{0.4cm}
\includegraphics[scale = 0.47, trim = {1cm 4cm 0cm 3cm}, clip]{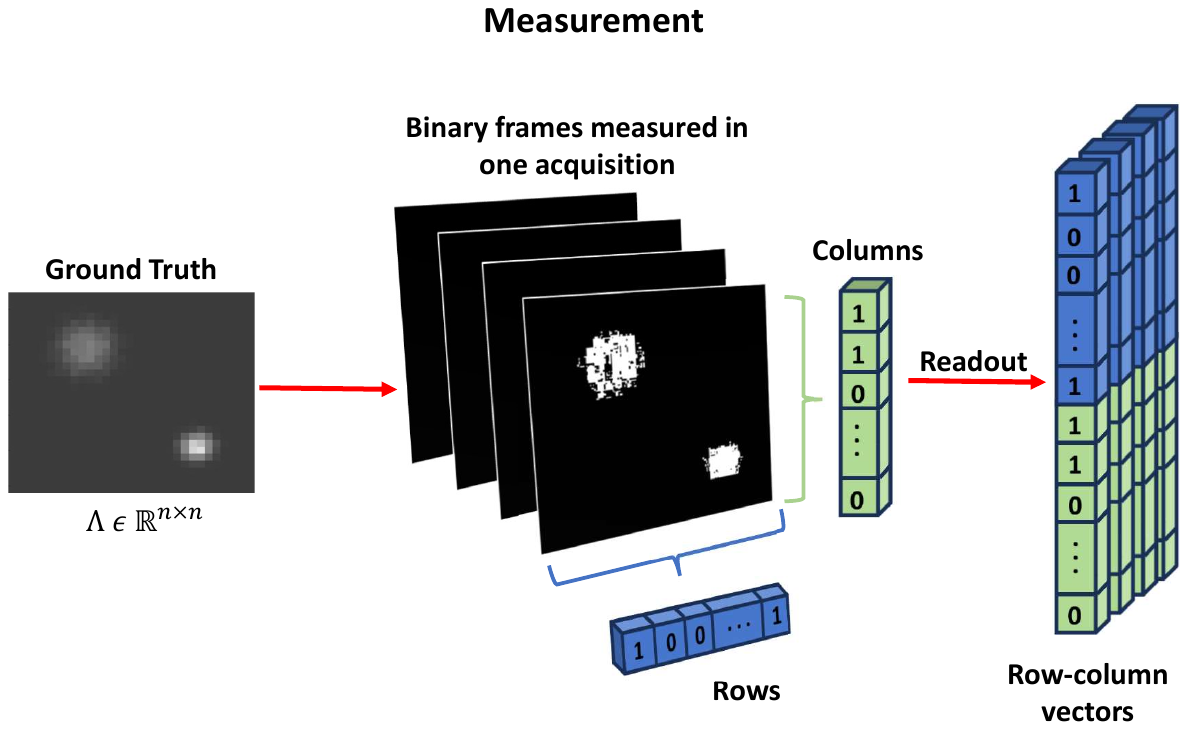}} \hspace{-2.3cm}
{
\includegraphics[scale = 0.47, trim = {6.1cm 4cm 3cm 3cm}, clip]{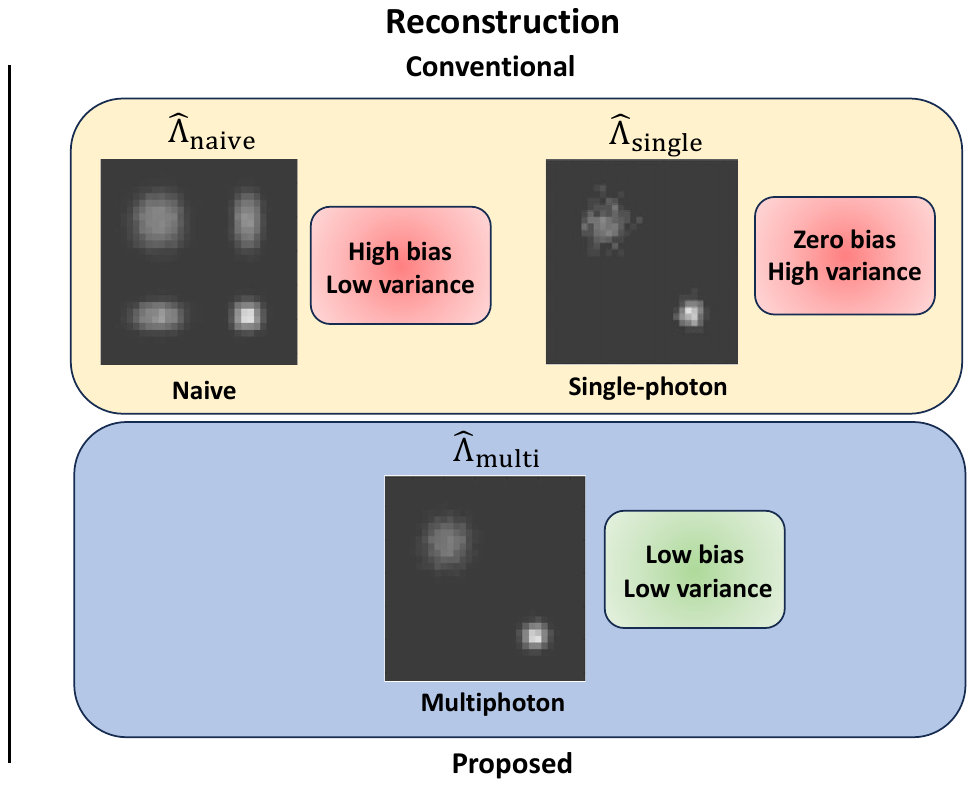}}
\caption{Schematic of image reconstruction with row--column readouts.
\emph{(Left)} Multiple frames of a ground truth image are measured using an SNSPD array to give row--column readout vectors.
\emph{(Right)} Comparison of conventional and proposed reconstruction techniques to estimate the ground truth.}

\label{fig:Concept}
\end{figure*}

\section{Measurement and Readout Model}
\label{sec:img_model}

\Cref{fig:Concept} illustrates imaging a scene using an $n\times n$ SNSPD array with row--column readouts. Measurements are read out after integrating the incident signal for a fixed time interval.
Let $\Lambda \in \mathbb{R}^{n \times n}$ be the flux of an $n \times n$ ground truth image in that integration time.
The number of photons arriving at pixel $(i,j)$ at a discrete time index $t$ is modeled as
\begin{equation}
X_{ij}^t \sim \rm{Poisson}(\Lambda_{\emph{ij}}).
\label{eq:forward_model_2}
\end{equation}
We assume pixel saturation with the detection of a single photon, i.e., the detectors are not photon-number resolving \cite{li2021saturation}.
Thus,
we define binary-valued photon-incidence indicator $Y^t \in \{0, 1\}^{n \times n}$
by
$Y^t_{ij} = \mathbf{1}[X^t_{ij} > 0]$,
where $\mathbf{1}[\cdot]$ is the indicator function.
Then, $Y^t$ is a matrix where each entry is an independent Bernoulli random variable $Y^t_{ij} \sim \text{Bernoulli}(q_{ij})$,
where
\begin{equation}
\label{eq:qij}
   q_{ij} = 1 - e^{-\Lambda_{ij}}
\end{equation} 
is the probability of photon detection at pixel $(i,j)$.

\begin{figure}
    \includegraphics[width=0.48\textwidth, trim = {1cm 3cm 1.5cm 3cm}, clip]{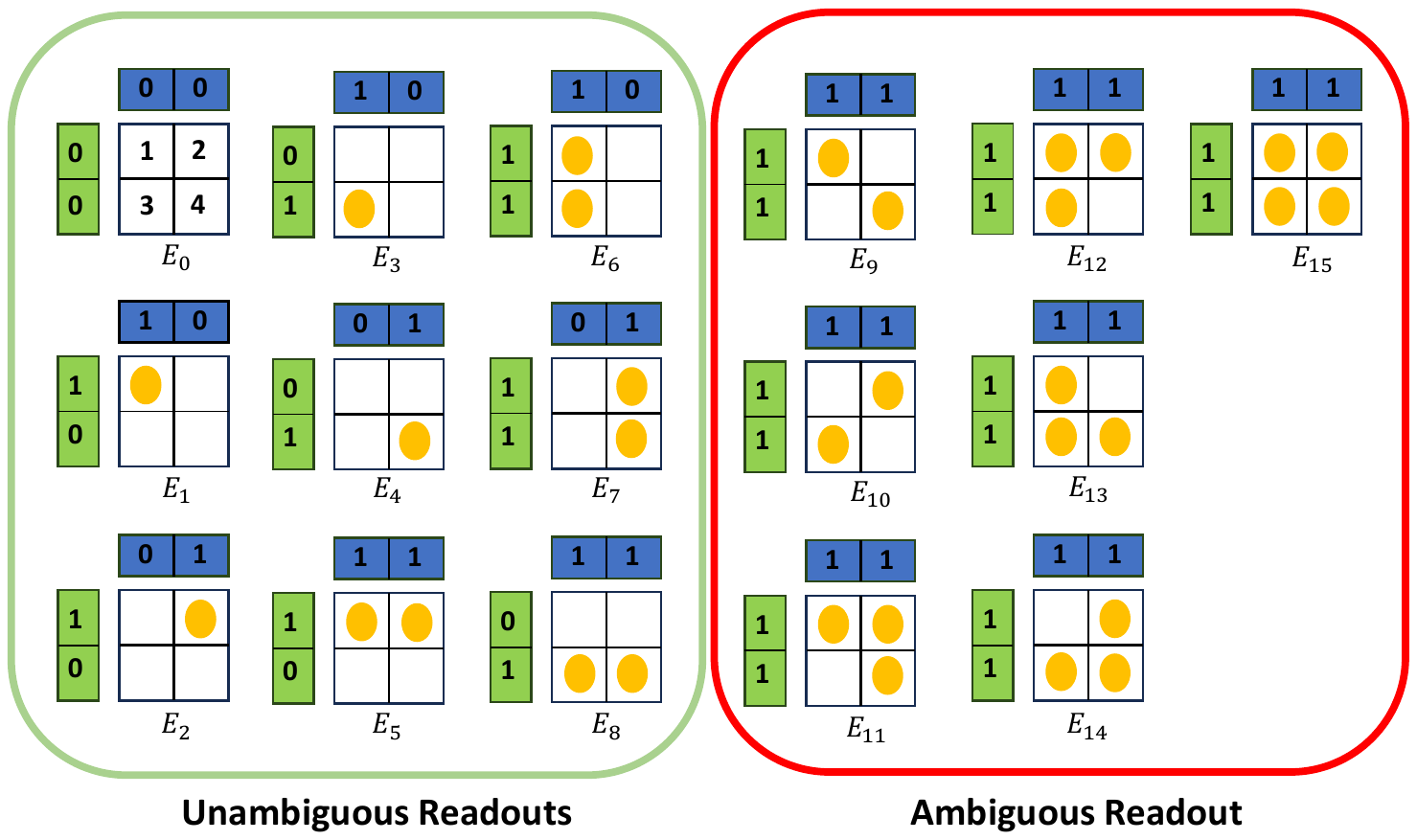}
    \caption{ Row--column readout frames for a $2 \times 2$ array.
    Yellow spots indicate incident photon locations.
    Accounting for pixel saturation, $16$ photon detection events are possible with this array.
    Row--column readout results in 9 unambiguous events $E_0,E_1,\ldots,E_8$ and 7 events $E_9,E_{10},\ldots,E_{15}$ that are ambiguous because they all result in $R^t = [1,1]$ and $C^t = [1,1]$. Pixel numbers are indicated in event $E_0$, which corresponds to an empty readout.}
    \label{fig:2x2 readouts}
\end{figure}

Each measurement consists of a row readout $R^t \in \{0, 1\}^n$ and a column readout $C^t \in \{0, 1\}^n$:
\begin{subequations}
\begin{align}
    &R^t_i = \mathbf{1}\!\left[ X^t_{ij} > 0 \text{ for any } j \in \{1, \ldots, n \}\right], \\
    &C^t_j = \mathbf{1}\!\left[ X^t_{ij} > 0 \text{ for any } i \in \{1, \ldots, n \}\right].
\end{align}
\end{subequations}
We then define a readout \emph{frame} at time $t$ as $(R^t, C^t)$. 
When multiple photons are incident on the array within a single integration period, the problem of resolving the spatial locations of their incidence is ill-posed. 
As an example, consider a $2 \times 2$ array with pixels numbered from 1 to 4 as shown in event $E_0$ of \cref{fig:2x2 readouts}.
The frame $([1, 1], [1, 1])$ could have resulted from photons detected at pixels $\{1,4\}$ or $\{2,3\}$ or any of: 
$\{1,2,3\}$,
$\{1,2,4\}$,
$\{1,3,4\}$,
$\{2,3,4\}$, or
$\{1,2,3,4\}$.
More generally, a readout frame is unambiguous if
\(\sum_{i=1}^{n} R_i^t = 1\)
or
\(\sum_{j=1}^{n} C_j^t = 1\);
otherwise, it is ambiguous. 
Our problem is to  estimate $\Lambda$ from measurements in $T$ frames $\{(R^t, C^t)\}_{t = 1}^T$.
In particular, we are interested in exploiting the information contained in ambiguous readout frames.

\section{Estimators}
\def\bA{{\boldsymbol A}}
\def\cL{{\mathcal L}}
\def\cY{{\mathcal Y}}
\def\ghat{\widehat{g}}
\label{section:estim}
In this section, we introduce the three estimators compared in this study, namely the naive, single-photon, and multiphoton estimators.
Each is formulated first as an estimator $\widehat{q}_{ij}$ for the probability $q_{ij}$ defined in \eqref{eq:qij}.
The incident flux at each pixel can then be estimated elementwise as
\begin{equation} 
    \label{eq:lambda}
     \widehat{\Lambda}_{ij} = -\log\!\left(1-e^{-\widehat{q}_{ij}}\right).
\end{equation}

\subsection{Naive Estimator}
The \emph{naive estimator} (NE) assumes that a photon is detected at each of the candidate pixels in ambiguous readout frames.
Thus, the probability of detecting a photon at pixel $(i,j)$ is estimated as
\begin{equation} \label{eq:expNaive}
    \PNaive = \frac{1}{T} \sum_{t = 1}^{T} R_i^t C_j^t.
\end{equation}
The naive estimator overestimates $q_{ij}$
because it misattributes photon counts in ambiguous frames to pixels where no photon incidence occurred. 
The extra counts imputed
by this scheme produce ghost spots in the reconstructed image, as depicted in \cref{fig:Concept}.\\
The positive bias of the naive estimator is 
\begin{align}
\mathbb{E}&{[\PNaive]} - q_{ij} \nonumber \\
  &= 
    (1 - q_{ij})
    \!\!\left(1-\prod_{k \in R_i'}(1-q_{kj})\right)
    \!\!\left(1-\prod_{\ell \in C_j'}(1-q_{i\ell})\right),
\label{eq:naive_bias}
\end{align}
where $R_i'$ is the set of rows except row $i$ and $C_j'$ is the set of columns except column $j$.
A derivation is given in the supplement.

\subsection{Single-Photon Estimator}
A simple way to prevent misattributions and thus reduce bias is to only use frames with a single detected photon in the reconstruction since these unambiguously give the spatial locations of photon incidence.
We thus define the \emph{single-photon estimator} (SPE): 
\begin{equation} \label{eq:singlePhotonEst}
    \PSingle = \begin{cases}
        \frac{N_{ij}}{N_{ij} + N_0}, &\text{ if } N_{ij} + N_0 > 0; \\
        0, &\text{ if } N_{ij} + N_0 = 0,
    \end{cases}
\end{equation}
where $N_{ij}$ is the number of frames where a photon is detected only at pixel $(i, j)$, and $N_0$ is the number of frames without any detected photons across the whole array.
The SPE is unbiased when $N_{i,j} + N_0 > 0$, as shown in the supplement.
However, the discarding of frames with multiple detected photons results in high variance.

\subsection{Multiphoton Estimator}
\label{subSection:MultiphotonEst}
Now, we develop a \emph{multiphoton estimator} (ME) that uses photon counts from ambiguous readouts for reconstruction.
For simplicity, here we derive expressions for detection probability estimates at each pixel of a $2 \times 2$ array and illustrate how this estimator maximizes an approximate likelihood of observations. The general form of the ME is presented in the supplement.
If each pixel were read out individually, the likelihood of observations could be written as
\begin{equation}
    \mathcal{L}_p = \prod_{i=1}^2\prod_{j=1}^2(1-q_{ij})^{v_{ij}}(q_{ij})^{f_{ij}},
    \label{eq:pixelwise_likelihood}
\end{equation}
where $v_{ij}$ is the total number of frames with no detections at pixel $(i,j)$ and $f_{ij}$ is the total number of frames with a detection at pixel $(i,j)$. Then, the maximum likelihood estimate $\widehat{q}_{ij}^{\rm ML}$ could be computed as 
\begin{equation}
    \widehat{q}_{ij}^{\rm ML} = \frac{f_{ij}}{T},
    \qquad
    \mbox{$i=1,2$, $j=1,2$},
\end{equation} 
where $T$ is the total number of measured frames.

With row--column readouts, our knowledge of locations of photon detections is reduced.
We define $v_{ij}$ and $f_{ij}$ from unambiguous frames only
and find that
the likelihood of observations becomes 
\begin{equation}
    \mathcal{L}(\{(R^t, C^t)\}_{t = 1}^T; q) = U(q)A(q) ,
    \label{eq:2x2Likelihood}
\end{equation}
with the factor
\begin{equation}
    U(q) = \prod_{i=1}^2\prod_{j=1}^2 (1 - q_{ij})^{v_{ij}} (q_{ij})^{f_{ij}}
    \label{eq:2x2_UnambTerm}
\end{equation}
from unambiguous frames
and the factor
\begin{equation}
\label{eq:simplified_A}
    A(q) = (q_{11}q_{22} + q_{12}q_{21} - q_{11}q_{12}q_{21}q_{22})^{M_9}
\end{equation}
from ambiguous frames.
Here, $M_9$ is the total number of ambiguous readout frames due to events $E_9, \ldots, E_{15}$ shown in \cref{fig:2x2 readouts}.

The factor in \eqref{eq:simplified_A} results from the sum of probabilities $\sum_{k=9}^{15} \mathbb{P}(E_k)$. This makes it challenging to find analytical expressions for the maximum likelihood estimates of detection probabilities. However, if we can estimate the fractions of the $M_9$ frames resulting from each of the events $E_9, \ldots, E_{15}$, we can re-write \eqref{eq:simplified_A} as a product of probabilities of events $E_9, \ldots, E_{15}$ to express \eqref{eq:2x2Likelihood} similarly to \eqref{eq:pixelwise_likelihood} and obtain analytical expressions for approximate maximum likelihood estimates $\widehat{q}_{ij}^{\rm a}$.
We first find expressions for the conditional probabilities $g_9, \ldots, g_{15}$ of events $E_9, \ldots, E_{15}$, given that an ambiguous frame was read out. For example, $g_9$ is the conditional probability of event $E_9$ given that a frame $R^t = [1, 1]$, $C^t = [1, 1]$ was measured and is given by
\begin{equation}
    {g}_{9} = \frac{\mathbb{P}(E_9)}{\sum_{k = 9}^{15} \mathbb{P}(E_k)}.
    \label{eq:g_9}
\end{equation}
In addition to the inherent limitation of having a finite amount of data, computing
$\mathbb{P}(E_9), \ldots, \mathbb{P}(E_{15})$ is hindered by ambiguities. Thus, we estimate these probabilities using the measured single-photon frames, which are unambiguous and unbiased. 
For example, an estimate of $\mathbb{P}(E_9)$ can be calculated as
\begin{equation}
   \widehat{\mathbb{P}}(E_9) = \widehat{q}_{11}^{\rm s}\widehat{q}_{22}^{\rm s}(1-\widehat{q}_{12}^{\rm s})(1-\widehat{q}_{21}^{\rm s}).
\end{equation}
Similar expressions can be derived for $\widehat{\mathbb{P}}(E_{10}), \ldots,\widehat{\mathbb{P}}(E_{15})$, which are then substituted in \eqref{eq:g_9} in place of $\mathbb{P}(E_9), \ldots ,\mathbb{P}(E_{15})$ to get estimates
$\ghat_9, \ldots, \ghat_{15}$
of conditional probabilities $g_9, \ldots, g_{15}$.
The fractions of the $M_9$ frames resulting from events $E_9, \ldots E_{15}$ can then be estimated as $\ghat_9M_9, \ldots, \ghat_{15}M_9$. Finally, these estimates can be used to simplify \eqref{eq:simplified_A} to a product of probabilities and obtain the approximate maximum likelihood estimates
\begin{equation} 
  \begin{aligned}
     \widehat{q}^{\rm a}_{11} &= \frac{M_1 + M_5 + M_6 + M_9(\ghat_9    + \ghat_{11} + \ghat_{12} + \ghat_{13} + \ghat_{15})}{T},\\
     \widehat{q}^{\rm a}_{12} &= \frac{M_2 + M_5 + M_7 + M_9(\ghat_{10} + \ghat_{11} + \ghat_{12} + \ghat_{14} + \ghat_{15})}{T},\\
     \widehat{q}^{\rm a}_{21} &= \frac{M_3 + M_6 + M_8 + M_9(\ghat_{10} + \ghat_{12} + \ghat_{13} + \ghat_{14} + \ghat_{15})}{T},\\
     \widehat{q}^{\rm a}_{22} &= \frac{M_4 + M_7 + M_8 + M_9(\ghat_{9}  + \ghat_{11} + \ghat_{13} + \ghat_{14} + \ghat_{15})}{T},
  \end{aligned}
\end{equation}
where $M_1, \ldots, M_8$ are the numbers of unambiguous readout frames corresponding to events $E_1, \ldots, E_8$.

While computing the bias of the ME analytically is complex, numerical simulations show that it has a low bias. Furthermore, since the ME depends on the SPE for an initial estimate of the detection probabilities, when the SPE is poor due to few measured single-photon frames, the reduction in MSE from the ME is relatively lesser than in cases when the SPE has low variance.

\subsection{Scaling to Higher Dimensional Arrays}
\label{subsection:scaling}
In a $2 \times 2$ array, there is a single type of ambiguous readout which indicates photon incidences at both rows and both columns of the array.
As the spatial dimension of the array increases, there are more types of ambiguous readouts. 

For instance, with a $3 \times 3$ array, we can have the following ambiguous readouts:
\begin{itemize}
    \item 2 rows and 2 columns fired -- 4 candidate pixel locations;
    \item 2 rows and 3 columns fired -- 6 candidate pixel locations;
    \item 3 rows and 2 columns fired -- 6 candidate pixel locations;
    \item 3 rows and 3 columns fired -- 9 candidate pixel locations.
\end{itemize}
Resolving these additional ambiguous readouts requires the modification of the expressions for conditional probabilities $g$. 
As the number of rows and columns with photon coincidences increases, the number of unique terms in the expressions for $g$ grows rapidly.
Hence, we restrict our analysis and simulations to at most 4-photon coincidences and ignore terms resulting from 5-photon coincidences and above.
We expect that this assumption introduces a small bias in the ME, especially at higher incident fluxes. Nevertheless, the improvement over handling only single-photon frames can be substantial.

\section{Results}
\subsection{Imaging Arbitrary Scenes}

\begin{figure}
\centering
\begin{minipage}[t]{0.44\textwidth}
    
    \vspace{0.8cm}
    \begin{overpic}[width=\textwidth, trim = {2.7cm 2cm 2.2cm 2cm}, clip]{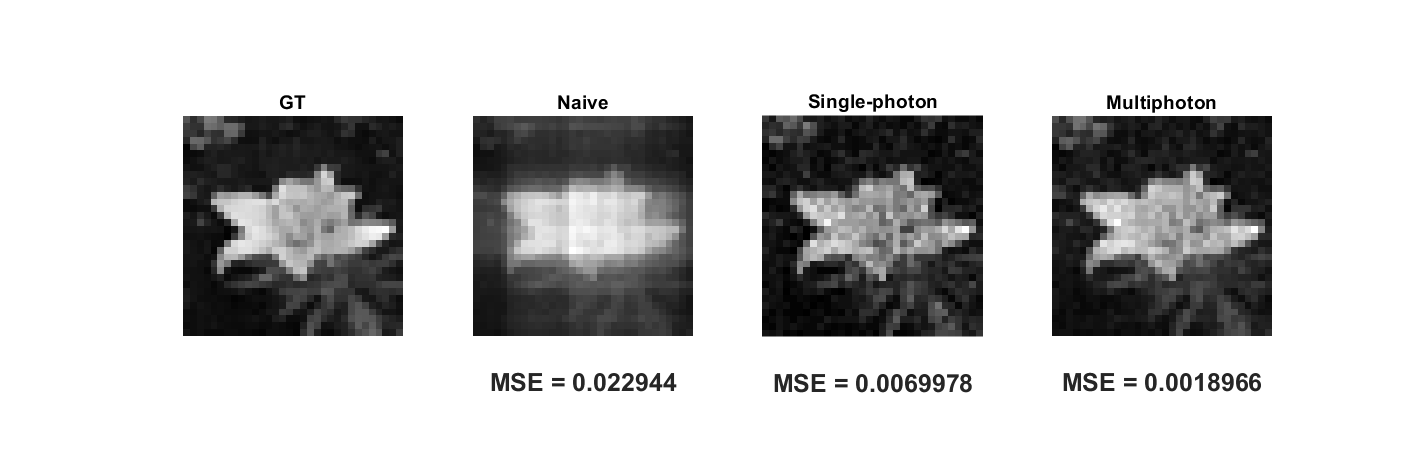}
        \put (-5, 10) {(a)}
        \put ( 0.5,   23) {\small Ground Truth}
        \put (32.5, 23) {\small Naive}
        \put (32.5, -3.5) {\fontsize{8pt}{12pt}\selectfont 16.39 dB}
        \put (52,   23) {\small Single-Photon}
        \put (58, -3.5) {\fontsize{8pt}{12pt}\selectfont 21.55 dB}
        \put (79.5, 23) {\small Multiphoton}
        \put (84, -3.5) {\fontsize{8pt}{12pt}\selectfont 27.22 dB}
    \end{overpic}
    \vspace{0.01cm}
    
    \begin{overpic}[width=\textwidth, trim = {2.7cm 2cm 2.2cm 2cm}, clip]{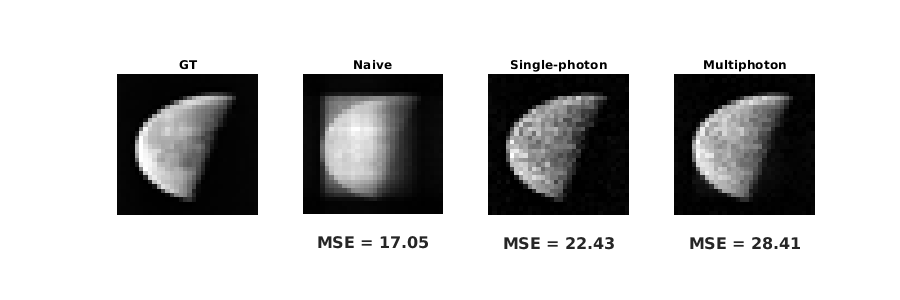}
        \put (-5, 10) {(b)}
        \put (32.5, -2.5) {\fontsize{8pt}{12pt}\selectfont  17.05 dB}
        \put (58, -2.5) {\fontsize{8pt}{12pt}\selectfont  22.43 dB}
        \put (84, -2.5) {\fontsize{8pt}{12pt}\selectfont  28.41 dB}
    \end{overpic}
    \vspace{0.01cm}
    
    \begin{overpic}[width=\textwidth, trim = {2.7cm 2cm 2.2cm 2cm}, clip]{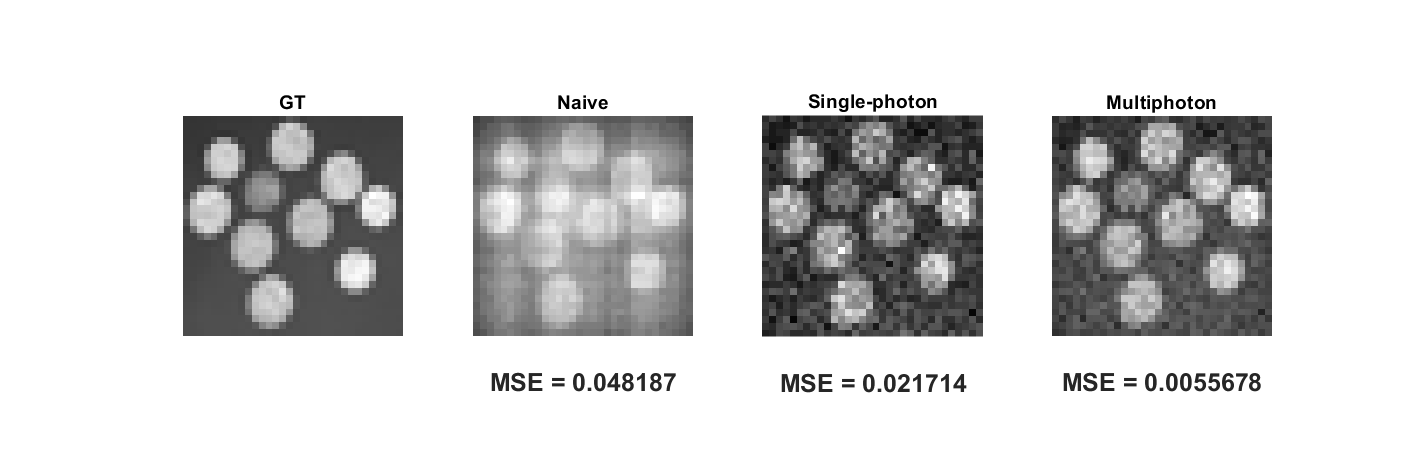}
        \put (-5, 10) {(c)}
        \put (32.5, -2.5) {\fontsize{8pt}{12pt}\selectfont  13.17 dB}
        \put (58, -2.5) {\fontsize{8pt}{12pt}\selectfont  16.63 dB}
        \put (84, -2.5) {\fontsize{8pt}{12pt}\selectfont  22.54 dB}
    \end{overpic}
    \vspace{0.01cm}
    
    \begin{overpic}[width=\textwidth, trim = {2.7cm 2cm 2.2cm 2cm}, clip]{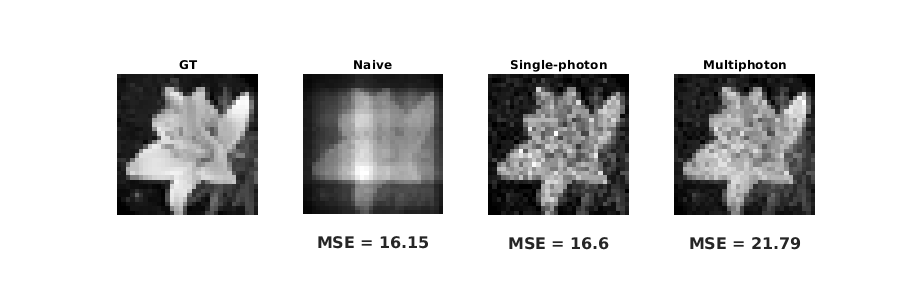}
        \put (-5, 10) {(d)}
        \put (32.5, -2.5) {\fontsize{8pt}{12pt}\selectfont 16.15 dB}
        \put (58, -2.5) {\fontsize{8pt}{12pt}\selectfont 16.60 dB}
        \put (84, -2.5) {\fontsize{8pt}{12pt}\selectfont 21.79 dB}
    \end{overpic}
    \vspace{0.01cm}
        
    \begin{overpic}[width=\textwidth, trim = {2.7cm 2cm 2.2cm 2cm}, clip]{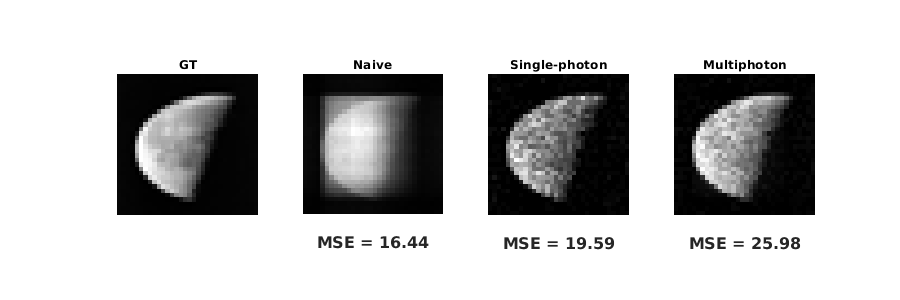}
        \put (-5, 10) {(e)}
        \put (32.5, -2.5) {\fontsize{8pt}{12pt}\selectfont  16.44 dB}
        \put (58, -2.5) {\fontsize{8pt}{12pt}\selectfont  19.59 dB}
        \put (84, -2.5) {\fontsize{8pt}{12pt}\selectfont  25.98 dB}
    \end{overpic}
    \vspace{0.01cm}
\end{minipage}
\vspace{-0.2cm}
\caption{Reconstruction of arbitrary $32 \times 32$ images from 100\,000 measured frames using the naive, single, and multiphoton estimators. Each PSNR value is from averaging over 100 trials.
Rows (a)--(c) show that the multiphoton estimator is 4 to 6~dB better than the single-photon estimator and around 6 to 11~dB better than the naive estimator. Row (d) shows the impact of rotating the ground truth image on the reconstruction. Row (e) shows the impact of increasing the average number of photons per frame from 3 to 4\@.
The SPE rejects $\sim$85\% of measured frames at a PPF of 3\@.
}
\label{fig:Natural Images}
\end{figure}

\Cref{fig:Natural Images} shows the Monte Carlo simulation results of reconstructing $32 \times 32$ images from 100,000 measured frames using each of the estimators.
Note that these are estimates of the flux $\widehat{\Lambda}_{ij}$ at each pixel, computed using the detection probability estimate $\widehat{q}_{ij}$ from the three estimators followed by the transformation in \eqref{eq:lambda}.

Reconstructions in rows (a), (b), (c), and (d) are at a scaling of 3 mean photons per frame (PPF),
while row (e) is at a scaling of 4 mean PPF\@.
Mean PPF is the average number of detected photons in each measured frame. 
Each experiment is repeated 100 times for the computation of PSNR\@.

In rows (a), (b), and (c), it can be seen that the NE introduces horizontal and vertical streaks due to misattributions from multiphoton coincidences. 
For example, in row (a) the SPE achieves $\sim$5 dB PSNR improvement over the NE and does not show the same artifacts, although the petals of the flower appear noisy.

Our proposed ME outperforms the naive estimator by $\sim$11dB while its improvement over the single-photon estimator is $\sim$6\,dB\@.
Further, we note that the ME preserves features in the ground truth image better than both of the other estimators, in addition to eliminating misattribution artifacts. The petals of the flower in row (a) are more clearly visible, while these are missing in the other reconstructions.
The improvement in the reconstruction depends on some features of the ground truth image. For example, in row (d) we consider a rotated version of the image in row (a). It can be seen that the misattributions in the NE are more concentrated towards the center-left of the image as compared to row (a) where the horizontal and vertical streaks cut across the image. The PSNR of the naive reconstruction is lower than that in row (a), however, the multiphoton reconstruction still achieves $\sim$5dB improvement.
In row (e) we consider the image from row (b) with a PPF scaling of 4\@.
We note that the PSNR of the NE reduces by $\sim$0.6 dB while the PSNR of the SPE and ME reduce by $\sim$3 dB each. This behavior is expected as an increase in the average PPF causes the number of frames with a single detected photon to decrease, which reduces the PSNR of the SPE and consequently the ME\@. However, the performance of the ME can be improved by considering additional multiphoton events.

\subsection{Optimal Photons Per Frame}
\begin{figure}
 \centering
  \centering
  \includegraphics[width = 0.5\textwidth]{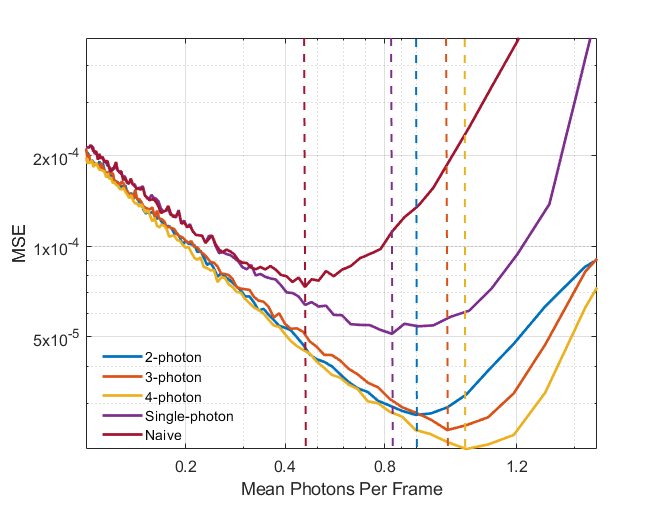} 
 \caption{Mean-squared errors of the estimators studied as functions of mean photons per frame.
 (Ground truth is image in \cref{fig:Natural Images}(c).)
 \label{fig:OptimPPF}
}
\end{figure}

In single-photon imaging, high signal strength is often beneficial for parameter estimation. However, for readout-multiplexed arrays, high flux also leads to more ambiguous readout frames, which makes image reconstruction challenging. We therefore expect that there exists an optimal incident photon flux level where each of the estimators achieves the minimum mean-square error. The following results, as shown in \cref{fig:OptimPPF}, validate this hypothesis.

A ground truth image can be scaled in experimental settings by selecting a desired integration time or attenuating the incident photon flux using a neutral density filter. To study the dependence of estimator performance on the incident photon flux, we perform a simulation with a high emitted photon flux from a source that is attenuated before being incident on the array. 

We then study the change in mean-squared error per pixel ($\textrm{MSE} = \normShort{\widehat{\Lambda} - \Lambda}_2^2 /n^2)$ of the reconstructed image with varying strength of the attenuating filter.
The results of this simulation are shown in \cref{fig:OptimPPF} for the reconstruction of a $32 \times 32$ ground truth image in row (c) of \cref{fig:Natural Images}.

Note that as the mean PPF increases, the MSEs of all the estimators initially decrease, reach a minimum, and then increase. 
High MSE at low PPF values is due to the lack of detected photons for accurate reconstruction. The increase in MSE at high PPF values is expected due to the high variance of SPE and the high probability of multiphoton events not modeled by the ME\@. 

It can also be seen that the minimum MSE of the multiphoton estimator is lower than that of the single-photon estimator, and both of these are much lower than the naive estimator. Furthermore, the minimum value of MSE for the ME occurs at a PPF of 1.4, which is higher than that for the SPE at 0.83, and that for NE at 0.45\@. We expect these incidence flux values to be realistic in conditions similar to imaging under moonlight, where high-efficiency detectors like SNSPDs are typically used. Thus, the ME produces a better reconstruction compared to both the NE and the SPE when multiple photons are detected simultaneously at the array. This results in an increased photon count rate and a better utilization of the spatial dimensions of the array. 

\subsection{2-, 3- and 4-Photon Estimators}
As explained in \cref{subsection:scaling}, our estimator resolves up to 4 simultaneously incident photons. The conditional probability expressions ($g$'s)
contain terms corresponding to 
every possible photon event that can result in the measured readout. 

However, we can simplify these expressions by restricting the number of modeled events. For instance, if we include only 2-photon coincidences in the case of a $2 \times 2$ array, only conditional probabilities of events $E_9$ and $E_{10}$ would be summed in \eqref{eq:g_9}.
\cref{fig:OptimPPF} shows the performance of the ME when 2-, 3- and 4-photons are modeled in $g$. It can be seen that the MSE of reconstruction is successively reduced with each additional photon considered in the estimator. The 2-photon estimator achieves its lowest MSE at a mean PPF of 1, while the 3- and 4-photon estimators achieve their lowest MSEs at 1.2 and 1.4 mean PPF, respectively. 

\begin{figure}
    \vspace{0.5cm}
    \begin{overpic}[width=0.5\textwidth, trim = {8cm 10cm 8cm 8cm}, clip]{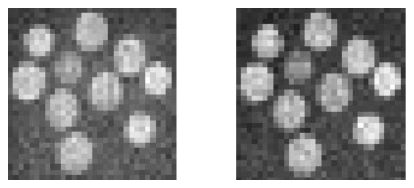}
        
        \put ( 30,   28) {\fontsize{10pt}{12pt}\selectfont Naive}
        \put ( 29,   -5) {\fontsize{10pt}{12pt}\selectfont 22.59 dB}
        \put (56, 28) {\fontsize{10pt}{12pt}\selectfont Single-Photon}
        \put (61,   -5) {\fontsize{10pt}{12pt}\selectfont 23.64 dB}

    \end{overpic}
    \vspace{0.6cm}
    
    \begin{overpic}[width=0.48\textwidth, trim = {9cm 10cm 8cm 8cm}, clip]{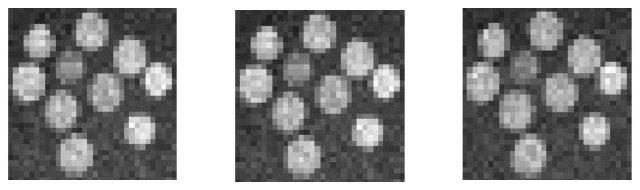}
        \put (5,   28) {\fontsize{10pt}{12pt}\selectfont 2-photon}
        \put (6,   -6) {\fontsize{10pt}{12pt}\selectfont 23.83 dB}
        \put (39, 28) {\fontsize{10pt}{12pt}\selectfont 3-photon}
        \put (40, -6) {\fontsize{10pt}{12pt}\selectfont 24.60 dB}
        \put (75, 28) {\fontsize{10pt}{12pt}\selectfont 4-photon}
        \put (76, -6) {\fontsize{10pt}{12pt}\selectfont 26.74 dB}
    \end{overpic}
    \vspace{0.3cm}
    \caption{Image reconstruction by operating each of the studied estimators at mean PPF values corresponding to their lowest MSE's. Indicated values are PSNR of reconstructions.}
    \label{fig:Coins2Rows}
\end{figure}

Image reconstruction using each of the estimators in \cref{fig:OptimPPF} at their optimal mean PPF values is summarized in \cref{fig:Coins2Rows}.
We see that the 4-photon estimator achieves the highest PSNR, which is 3 to 4 dB higher than the NE and the SPE. 

\subsection{Reduction in Total Integration Time}
\begin{figure}
    \begin{minipage}[t]{0.48\textwidth}
        \hspace{-0.2cm}
        \includegraphics[width=\linewidth, trim={0cm 0cm 1cm 0cm}, clip]{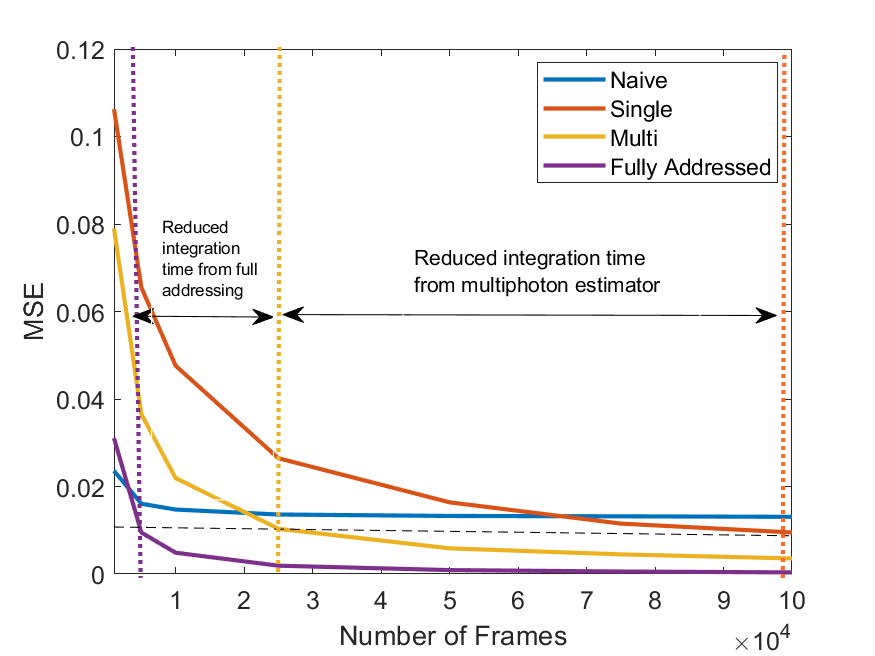}
        \caption{Reduction in the required number of frames to achieve the same target MSE by the NE, SPE,ME, and a fully-addressed readout scheme.}
        \label{fig:Frames_Reduced}
    \end{minipage}
\end{figure}

Since the multiphoton estimator enables the accurate interpretation of photon coincidences, it can be leveraged to reduce integration times in imaging with single-photon detectors. The performance of the three estimators in achieving a target reconstruction MSE is shown in \cref{fig:Frames_Reduced}. 
It can be seen that to achieve an MSE of 0.01, the ME requires $25,000$ measured frames, and the SPE requires $100,000$ measured frames, while the NE does not achieve this MSE within the range of frame values studied.
This shows that the ME achieves a factor $\sim$4 reduction in the integration time for the same reconstruction quality as the other estimators. In comparison, a fully-addressed readout scheme achieves the target MSE with only $5000$ frames, but with the increased complexity of readout.

\subsection{Cram{\'e}r--Rao Bound}
The Cram{\'e}r--Rao bound (CRB) provides the lowest possible variance of an unbiased estimator for parameter estimation. Here, we compare the performance of the NE, SPE, 2-photon, 3-photon, and 4-photon estimators on the estimation of a $2 \times 2$ ground truth, against CRB values at different incident photon fluxes. We first compute the Fisher information matrix (FIM) using the likelihood expression for row--column readouts from a $2 \times 2$ array. The complete derivation of expressions for elements of the FIM are shown in the supplement. The element $(1,1)$ of the FIM for a single measurement is

\begin{align}
    &\cI_{11} = \frac{(1-q_4)(1-q_2q_3)}{q_1} + \frac{(1-q_2q_3)}{1-q_1} \nonumber \\
    &\hspace{1cm}+ \frac{q_4^2(1-q_2q_3)^2}{q_1q_4 + q_2q_3 - q_1q_2q_3q_4}.
\label{eq:FI_terms}    
\end{align}

\begin{figure}
    \vspace{0.2cm}
    \begin{minipage}[b]{0.48\textwidth}
    \centering
        \includegraphics[width = \textwidth]{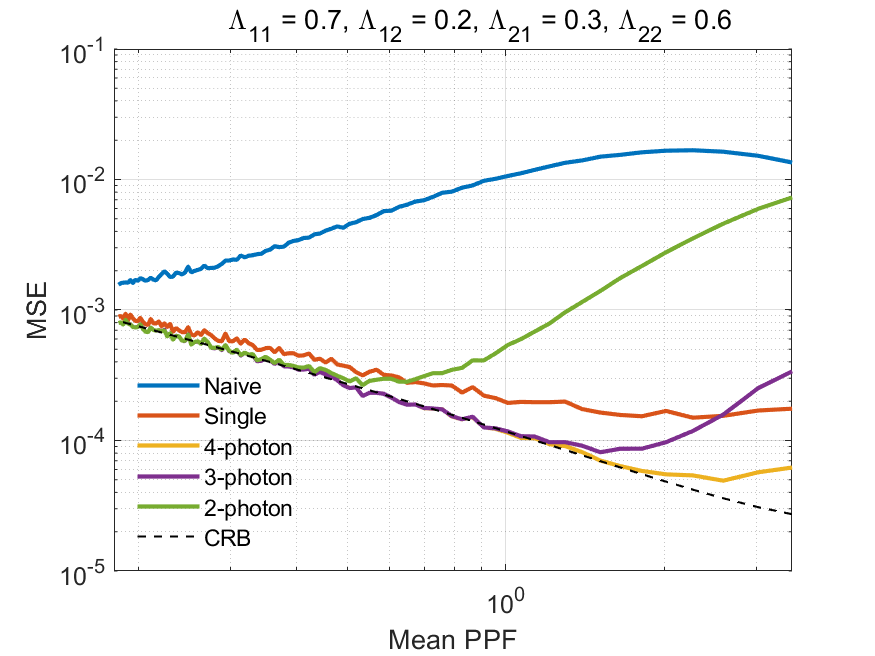}
    \caption{Comparison of estimator performance with the Cram{\'e}r--Rao Bound. 
    The ME matches the CRB for a range of incident photon fluxes and deviates from it at high fluxes. With 4 bright spots, model mismatch causes 2- and 3-photon estimators to become biased and deviate from the CRB at lower incident photon fluxes compared to the 4-photon estimator.
    }
    \label{fig:CRB}
    \end{minipage}
\end{figure}

Other diagonal elements have similar forms as shown in the supplement. The CRB is then computed as the mean of the diagonal elements in $\cI^{-1}$. The variation of CRB values, as well as MSEs of the studied estimators as a function of incident photon flux, is shown in \cref{fig:CRB}.

It can be seen that the ME curve matches the CRB curve for a range of mean PPF values. At high fluxes, the MSE of the ME increases due to the high variance of the single-photon estimates, which are used to calculate $g$ in \eqref{eq:g_9}. In the general case with bright spots at all four pixels of the array, the 4-photon estimator most closely matches the CRB across the range of PPF values studied. The 2- and 3-photon estimators deviate from the CRB due to the modeling of only a subset of events that can cause ambiguities and omission of terms in the expression for conditional probabilities in \eqref{eq:g_9}. The dependence of the estimator performance on the ground truth image is shown in the Supplement.

\subsection{Increasing array resolution}
\label{subSection:ArraySize}
Our analysis of the multiphoton estimator's performance so far has been limited to $32 \times 32$ arrays to match the sensor developed in ~\cite{wollman2019kilopixel}. 
Here, we study its use in arrays with higher pixel counts and finer spatial resolutions. We emulate a fixed sensor area with a varying number of pixels and fixed maximum photon flux. This results in photons per pixel being inversely proportional to the number of pixels. We include attenuation optimization for each reconstruction method, similar to \cref{fig:OptimPPF}.

\begin{figure}
 
    \centering
    
    \includegraphics[width=\linewidth, trim = {0cm 0.6cm 0cm 0cm}, clip]{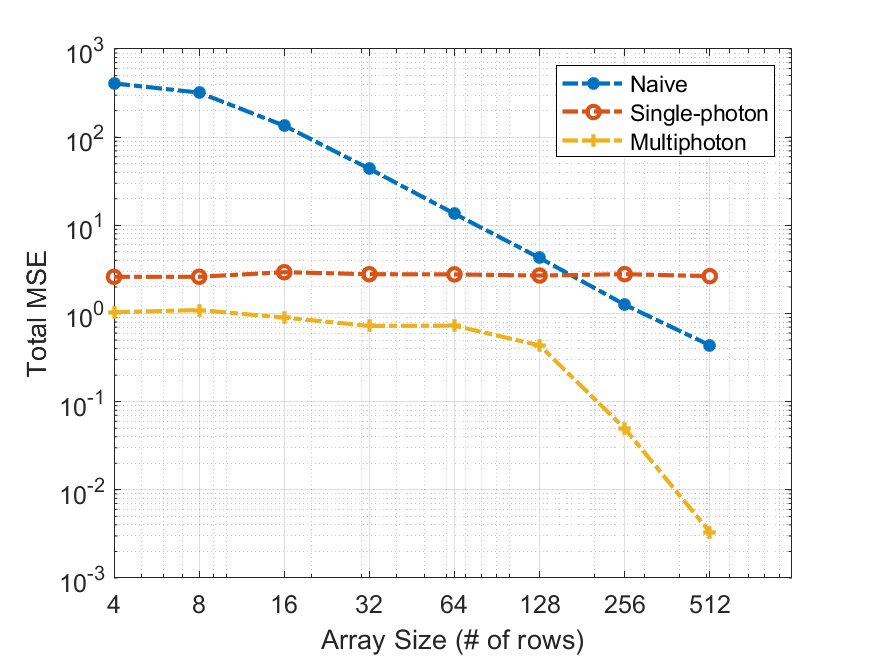} 
    {\footnotesize (a) MSE comparison} 
    \includegraphics[width=\linewidth]{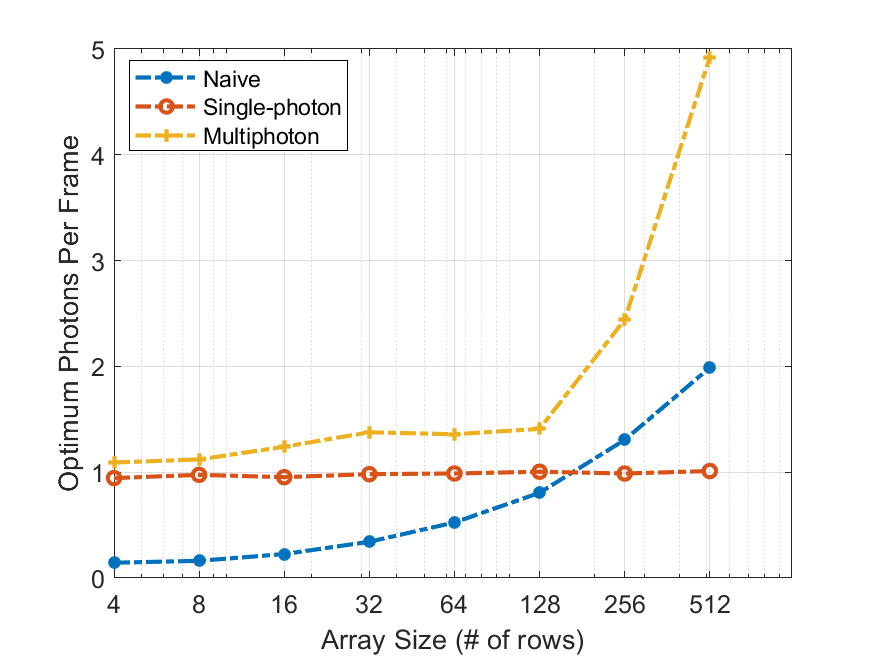} \\
    {\footnotesize (b) Optimal PPF comparison} \\[-2mm]
    \caption{
    Influence of array size.
    (a) MSEs at optimum attenuation for each estimator.
    (b) Optimum photons per frame for each estimator.
    }
     \label{fig:ArraySize}
    \vspace{-4mm}
\end{figure}

Results for row 1 of \cref{fig:Natural Images} as the ground truth
are shown in \cref{fig:ArraySize}
Performance improvement from the ME increases with array size (see \cref{fig:ArraySize}(a)).
This shows that our method is applicable to array sizes encountered in commercial imaging use cases. Nevertheless, we expect coincidences not modeled to be a limiting factor. 
Further, computation time increases linearly with an increase in array size. Thus, for a practical implementation of our algorithm, it could be beneficial to divide the sensor area into smaller blocks, the readouts from each of which can be processed using the ME\@.
Such a processing approach would be particularly beneficial in sensor architectures similar to~\cite{hao2024compact}.
\section{Conclusions and Future Works}
\vspace{-2mm}
We formulate multiphoton coincidence resolution in readout multiplexed single-photon detector arrays as an inverse imaging problem and propose a novel estimator to resolve up to four photon coincidences with row--column multiplexed readouts. Our method enables high-flux, low-latency, high-resolution image reconstructions that significantly improve upon conventional methods. 
Future work can extend this method to resolve an abitrarily large number of coincident photon events or other types of multiplexed readouts. The use of spatial priors and deep learning methods can be explored to improve the presented solutions. This line of work could be especially useful for the use of SNSPD arrays in photon-starved imaging applications like deep space imaging or biological imaging.

\section*{Acknowledgements}
\label{sec:acknowledgements}
This work was supported in part by the U.S. National Science Foundation under Grant 2039762 and in part by a gift from Dr.\ John Zheng Sun. We thank Prof.\ Hoover Rueda-Chac{\'o}n for helpful discussions.

{
    \small
    \bibliographystyle{ieeenat_fullname}
    \bibliography{references,bibl}
}

 \clearpage
\setcounter{page}{1}
\setcounter{section}{0}
\maketitlesupplementary
\def\LambdaNaive {\widehat{\Lambda}_{ij, \rm naive}}
\def\LambdaSingle{\widehat{\Lambda}_{\rm single}}
\def\LambdaMulti {\widehat{\Lambda}_{\rm multi}}
\def\PNaive {\widehat{q}_{ij, \rm naive}}
\def\PSingle{\widehat{q}_{ij, \rm single}}
\def\PMulti {\widehat{q}_{ij, \rm multi}}
\newcommand{\ex}[1]{\mathbb{E}\left[{#1}\right]}
\newcommand{\Bin}{\text{Binomial}}

\section{Bias Calculations}
\label{sec:BiasCalc}
As noted in \cref{fig:Concept} and \cref{section:estim} of the main paper, the single-photon estimator is unbiased and the naive estimator has a positive bias. Here, we analytically derive these results.
\subsection{Naive Estimator}
For a given readout frame $(R^t,C^t)$, the naive estimator imputes a photon count to pixel $(i,j)$ when there is a detection at $(i,j)$ (with probability $q_{ij}$) or when detections occur in row $i$ and column $j$, but not at pixel $(i, j)$. The probability of the latter is
    \[
    (1 - q_{ij})
    \left(1-\prod_{k \in R_ i'}(1-q_{kj})\right)
    \left(1-\prod_{\ell \in C_ j'}(1-q_{i\ell})\right),
    \]
where $R_i'$ and $C_j'$ are the sets of rows and columns excluding row $i$ and column $j$, respectively.
Then $R_i^t C_j^t \sim \text{Bernoulli} (q'_{ij})$, where 
\begin{align}
q'_{ij}
  = & q_{ij}
  + (1 - q_{ij}) \nonumber \\
  & \left(1 - \prod_{k \in R_i'}(1-q_{kj})\right)
   \left(1 - \prod_{\ell \in C_j'}(1-q_{i\ell})\right).
  \label{eq:qij'}
\end{align}
Therefore, the bias of the naive estimator is
\begin{align}
\mathbb{E}&{[\PNaive]} - q_{ij} \nonumber \\
  &= 
    (1 - q_{ij})
    \!\left(1-\prod_{k \in R_i'}(1-q_{kj})\right)
    \!\left(1-\prod_{\ell \in C_j'}(1-q_{i\ell})\right).
\label{eq:naive_bias_suppl}
\end{align}

\subsection{Single-Photon Estimator}
As defined in \eqref{eq:singlePhotonEst}, the single-photon estimator is
${N_{ij}}/{(N_{ij} + N_0)}$ if $N_{ij} + N_0 > 0$ and 0 otherwise.
Here, $N_{ij}$ is the number of frames with a single detected photon at pixel $(i,j)$, and $N_0$ is the number of frames with no detected photons at the array.
We show this estimator to be unbiased when $N_{ij}+N_0 > 0$.
The expected value of the estimate is
\begin{align*}
    \ex{\PSingle} &\stackrel{(a)}= \ex{ \ex{\frac{N_{ij}}{N_{ij} + N_0} \Big| N_{ij} + N_0 } } \\
    &= \ex{ \frac{ \ex{N_{ij} | N_{ij} + N_0} }{N_{ij} + N_0} } \\
    &\stackrel{(b)}{=} \ex{ \frac{q_{ij} (N_{ij} + N_0) }{N_{ij} + N_0} } \\
    &= q_{ij},
\end{align*}
where (a) follows from the law of iterated expectation and (b) follows from the fact that
$N_{ij} \,|\, (N_{ij} + N_0) \sim \Bin ( R_{ij} / (R_0 + R_{ij}), N_{ij} + N_0 )$
with 
\begin{align*}
    R_0 &= \prod_{k = 1}^n \prod_{\ell = 1}^n (1 - q_{k\ell}), \\
    R_{ij} &= q_{ij} \prod_{k = 1, \ell = 1, (k,\ell) \neq (i,j)}^{k = n, \ell = n}(1 - q_{k\ell}).
\end{align*}

\section{General Form of the Multiphoton Estimator}
\label{sec:multiphoton}
The likelihood of all photon incidence frames $\{Y^t\}_{t = 1}^T$ is
\begin{equation}
    \cL(q; \{Y^t\}_{t = 1}^T) = \prod_{t = 1}^T p_Y(Y^t; q), \label{eq:indiLikelihood}
\end{equation}
where
\begin{equation}
    p_Y(y; q) = \prod_{i = 1}^m \prod_{j = 1}^n q_{ij}^{y_{ij}} (1 - q_{ij})^{1 - y_{ij}}
\end{equation}
is the probability mass function (PMF) of a photon incidence indicator. If we observe $\{Y^t\}_{t = 1}^T$, then the maximum likelihood estimator of $q$ is
\begin{equation}
    \widehat{q}_{ij} = \frac{1}{T} \sum_{t = 1}^T Y_{ij}^t. \label{eq:estFromY}
\end{equation}
With row--column readouts $\{(R^t, C^t)\}_{t = 1}^T$, a frame may be ambiguous, i.e., a readout $(R^t, C^t)$ can arise from many possible photon incidence events. Let $\bA: (R^t, C^t) \mapsto \cY^t$ be a mapping from a row--column readout to a set of possible photon incidence indicators. For example, in a $2 \times 2$ detector array,
\begin{equation}
   \bA\left(\begin{bmatrix} 1 \\ 1 \end{bmatrix}, 
   \begin{bmatrix} 1 \\ 1 \end{bmatrix}\right) = 
   \left\{
   \begin{array}{ccc}
       \begin{bmatrix} 1 & 0 \\ 0 & 1 \end{bmatrix},&
       \begin{bmatrix} 0 & 1 \\ 1 & 0 \end{bmatrix},&
       \begin{bmatrix} 1 & 1 \\ 0 & 1 \end{bmatrix},\\[2.5ex]
       \begin{bmatrix} 1 & 1 \\ 1 & 0 \end{bmatrix},&
       \begin{bmatrix} 1 & 0 \\ 1 & 1 \end{bmatrix},&
       \begin{bmatrix} 0 & 1 \\ 1 & 1 \end{bmatrix},\\[2.5ex]
       &\begin{bmatrix} 1 & 1 \\ 1 & 1 \end{bmatrix}\phantom{,}&
   \end{array}
   \right\},
\end{equation}
as illustrated in \cref{fig:2x2 readouts}. The probability of a readout $(R^t, C^t)$ is the sum of the probabilities of all possible photon incidence indicators. The PMF of a readout is
\begin{equation}
    p_{R, C}(r, c; q) = \sum_{Y \in \bA(r, c)} p_Y(Y; q).
\end{equation}
The likelihood of all row--column readout frames is
\begin{equation}
\label{eq:rcLikelihood}
    \cL(q; \{(R^t, C^t)\}_{t = 1}^T) = \prod_{t = 1}^T p_{R, C}(R^t, C^t; q) .
\end{equation}
Maximizing the likelihood \eqref{eq:rcLikelihood} is computationally difficult because the ambiguous frames render the log likelihood nonconcave with respect to $q$.
Instead of maximizing the photon incidence likelihood \eqref{eq:rcLikelihood}, the ME maximizes its approximation by distributing each ambiguous readout to possible photon incidence events according to $\widehat{q}_{\rm single}$. Let the approximate likelihood be
\begin{equation}
\label{eq:approxLikelihood}
\widetilde{\cL}(q; \{(R^t, C^t)\}_{t = 1}^T) = \prod_{t = 1}^T \widetilde{p}_{R, C}(R^t, C^t; q),
\end{equation}
where
\begin{align}
\widetilde{p}_{R, C}(r, c; q) = \prod_{Y \in \bA(r, c)} p_Y(g(Y, \bA(r, c)) Y; q)
\end{align}
and
\begin{equation}
\label{eq:condProb}
g(Y, \bA(r, c)) = \frac{p_Y(Y; \widehat{q}_{\rm single})}{\sum_{Y' \in \bA(r, c)} p_Y(Y'; \widehat{q}_{\rm single})}
\end{equation}
approximates the probability of $Y$ given a row--column readout $(r, c)$. 
For example, in a $2 \times 2$ array as demonstrated in \cref{fig:2x2 readouts}, there is only one type of ambiguous readout with $R^t = [1, 1]$ and $C^t = [1, 1]$ corresponding to events $E_9, \ldots, E_{15}$. 

Maximizing the approximate likelihood \eqref{eq:approxLikelihood} becomes similar to estimating $q$ from the photon incidence indicators $\{Y^t\}_{t = 1}^T$ as in \eqref{eq:estFromY}. The ME is therefore
\begin{equation}
    \widehat{q}_{ij, \rm{multi}} = \frac{1}{T} \sum_{t = 1}^T \sum_{Y \in \bA(R^t, C^t)} g(Y, \bA(R^t, C^t)) Y_{ij}.
\end{equation}
Intuitively, the ME estimates $q$ from a dataset of photon incidence indicators $\cup_{t = 1}^T \bA(R^t, C^t)$ synthesized from possibly ambiguous row--column readouts $\{(R^t, C^t)\}_{t = 1}^T$. In the estimator, each synthesized photon incidence indicator $Y \in \bA(R^t, C^t)$ is weighted according to the probability that it arises from the readout $(R^t, C^t)$ according to a preliminary estimate $\widehat{q}_{\rm single}$.

\section{Bias--Variance Decomposition of MSE}

\begin{figure}
    \centering
    \begin{tabular}{@{}c@{}}
    \includegraphics[width=\linewidth]{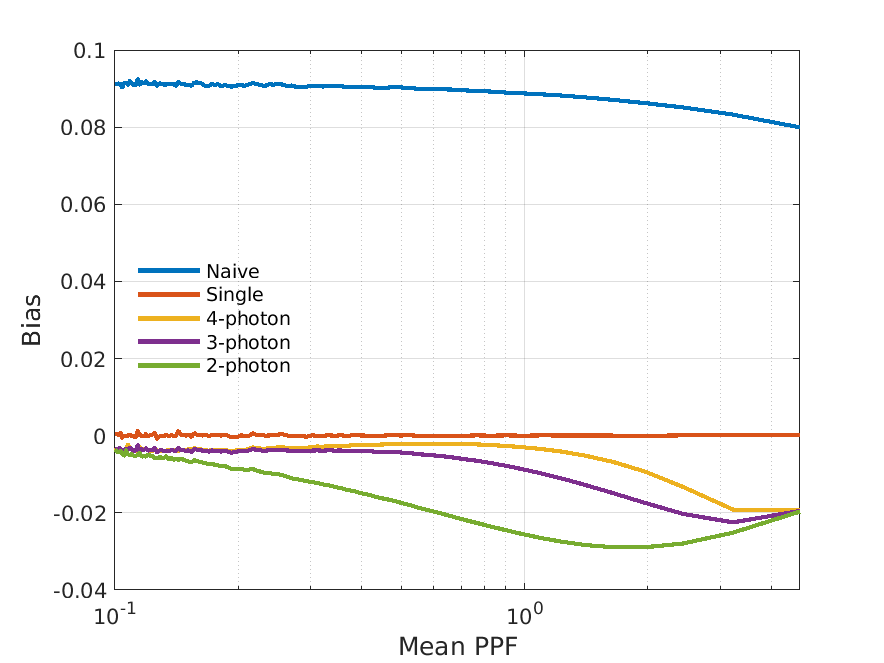} \\
    {\footnotesize (a) Bias} \\
    \includegraphics[width=\linewidth]{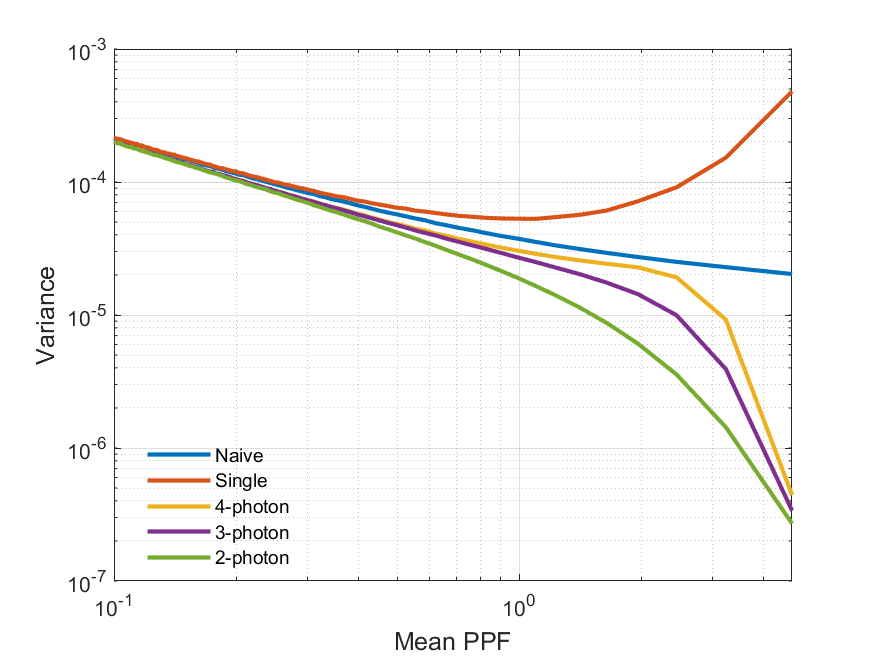} \\
    {\footnotesize (b) Variance} 
    \end{tabular}
    \caption{Change in bias and variance of the naive, single-photon, and multiphoton estimators as functions of mean photons per frame.}
    \label{fig:bias-var-decomp}
\end{figure}

\cref{fig:OptimPPF} shows the MSE as a function of mean PPF of the naive, single-, and multiphoton estimators. To better understand their behaviors, here we study the explicit change of bias and variance of each of the estimators as functions of mean PPF\@. 
\cref{fig:bias-var-decomp}(a) shows that the bias of the SPE is 0 across the range of PPF values, while the naive estimator has a large positive bias. The multiphoton estimator has a small negative bias that decreases as the number of photons modeled by the estimator increases. This negative bias is due to the multiphoton estimator imputing at most 4 coincident photons to ambiguous readouts that may have been produced by more coincident photons. 
\cref{fig:bias-var-decomp}(b) shows that the variance of the SPE is high, especially at high incident photon fluxes. This is expected since the SPE discards all ambiguous readouts, which have high probabilities of occurrence in the high incident flux cases. The ME achieves the lowest variance of the three estimators.

\section{Fisher Information Matrix Calculation}
The Cram{\'e}r--Rao bound in \cref{fig:CRB} is obtained by averaging over the diagonal elements of the inverse of the Fisher information matrix, which is derived using the log-likelihood expression for row--column readouts.
Here, we provide an explicit derivation of the entries in the FIM for a $2 \times 2$ array.

The number of readout frames of each type $M_0, \ldots, M_9$ can be modeled as
\begin{equation} 
    (M_0, M_1, \ldots, M_9) \sim  {\rm Multinomial}(r_0, r_1, \ldots, r_9, T),
\end{equation}
where $T$ is the number of measured frames and $r_0, \ldots, r_9$ are the probabilities of events $E_0, \ldots, E_9$ given by
\begin{equation}
    \begin{aligned}
        &r_0 = (1-q_{11})(1-q_{12})(1-q_{21})(1-q_{22}), \\
        &r_1 = (q_{11})(1-q_{12})(1-q_{21})(1-q_{22}), \\ 
        &r_2 = (q_{12})(1-q_{11})(1-q_{21})(1-q_{22}), \quad \ldots, \\
        &r_9 = (q_{11}q_{22} + q_{12}q_{21} - q_{11}q_{12}q_{21}q_{22}).
    \end{aligned}
    \label{eq:MultinomialProb}
\end{equation}

Further, for a $2 \times 2$ array, the expressions for $v_{ij}$ and $f_{ij}$ in the likelihood expression \eqref{eq:2x2Likelihood} are
\begin{equation}
\label{eq:vij_fij}
    \begin{aligned}
        v_{11} &= M_0 + M_2 + M_3 + M_4 + M_7 + M_8, \\ 
        f_{11} &= M_1 + M_5 + M_6, \\ 
        v_{12} &= M_0 + M_1 + M_3 + M_4 + M_6 + M_8, \\ 
        f_{12} &= M_2 + M_5 + M_7, \\
        v_{21} &= M_0 + M_1 + M_2 + M_4 + M_5 + M_7, \\
        f_{21} &= M_3 + M_6 + M_8, \\
        v_{22} &= M_0 + M_1 + M_2 + M_3 + M_5 + M_6,\\
        f_{22} &= M_4 + M_7 + M_8. \\
    \end{aligned}
\end{equation}
The Fisher information matrix of size $4\times 4$ is then calculated as 
\begin{equation} \label{eq:fiSecondDeriv}
    \cI = - \E \begin{bmatrix}
        \frac{\partial^2 \mathcal{L'}}{ \partial q_{11}^2} & \frac{\partial^2 \mathcal{L'}}{ \partial q_{11} \partial q_{12}} & \frac{\partial^2 \mathcal{L'}}{ \partial q_{11} \partial q_{21}} & 
        \frac{\partial^2 \mathcal{L'}}{ \partial q_{11} \partial q_{22}}\\
        \frac{\partial^2 \mathcal{L'}}{ \partial q_{12} \partial q_{11}} & \frac{\partial^2 \mathcal{L'}}{\partial q_{12}^2} & \frac{\partial^2 \mathcal{L'}}{ \partial q_{12} \partial q_{21}}
        & \frac{\partial^2 \mathcal{L'}}{ \partial q_{12} \partial q_{22}}\\
        \frac{\partial^2 \mathcal{L'}}{ \partial q_{21} \partial q_{11}} & \frac{\partial^2 \mathcal{L'}}{\partial q_{21} \partial q_{12}} & \frac{\partial^2 \mathcal{L'}}{\partial q_{21}^2}& \frac{\partial^2 \mathcal{L'}}{\partial q_{21} \partial q_{22}}\\
        \frac{\partial^2 \mathcal{L'}}{ \partial q_{22} \partial q_{11}} & \frac{\partial^2 \mathcal{L'}}{\partial q_{22} \partial q_{12}} & \frac{\partial^2 \mathcal{L'}}{\partial q_{22} \partial q_{21} }& \frac{\partial^2 \mathcal{L'}}{\partial q_{22}^2}
    \end{bmatrix},
\end{equation}
where
$\mathcal{L'} = \log(\mathcal{L})$.
Consider $\cI_{11}$.
First,
\begin{equation}
 \frac{\partial \mathcal{L'}}{ \partial q_{11}} =
    \frac{f_{11}}{q_{11}} - \frac{v_{11}}{1-q_{11}} + \frac{M_9q_{22}(1-q_{12}q_{21})}{q_{11}q_{22} +q_{12}q_{21} -q_{11}q_{12}q_{21}q_{22}} .
\end{equation}
Then,
\begin{align}
  \frac{\partial^2 \mathcal{L'}}{ \partial q_{11}^2} =& 
    -\frac{f_{11}}{q_{11}^2} - \frac{v_{11}}{(1-q_{11})^2} \nonumber \\
    &- \left(\frac{M_9q_{22}(1-q_{12}q_{21})}{q_{11}q_{22} +q_{12}q_{21} -q_{11}q_{12}q_{21}q_{22}}\right)^2 .
    \label{eq:SecondDerivLikelihood}
\end{align}
Since $M_0, \ldots, M_9$ are multinomial random variables, $\E[M_i] = Tr_i$. Thus, taking the negative expectation of \eqref{eq:SecondDerivLikelihood} and simplifying gives
\begin{subequations}
\begin{align}
    \cI_{11} =& \frac{(1-q_{22})(1-q_{12}q_{21})}{q_{11}} + \frac{(1-q_{12}q_{21})}{1-q_{11}} \nonumber \\
    &+ \frac{q_{22}^2(1-q_{12}q_{21})^2}{q_{11}q_{22} + q_{12}q_{21} - q_{11}q_{12}q_{21}q_{22}}.
\end{align}
Following a similar procedure, the remaining entries of the first row are
\begin{align}
      \cI_{12} &= \frac{q_{21}q_{22}}{q_{11}q_{22} + q_{12}q_{21} - q_{11}q_{12}q_{21}q_{22}}, \\
      \cI_{13} &= \frac{q_{12}q_{22}}{q_{11}q_{22} + q_{12}q_{21} - q_{11}q_{12}q_{21}q_{22}}, \\
      \cI_{14} &= \frac{q_{12}q_{21}(q_{12}q_{21}-1)}{q_{11}q_{22} + q_{12}q_{21} - q_{11}q_{12}q_{21}q_{22}}.
\end{align}
\end{subequations}
Similar expressions are derived for all the entries in the Fisher information matrix.

\begin{figure*}
    \centering
    \begin{tabular}{@{}c@{}c@{}c@{}}
        \includegraphics[width=0.33\textwidth]{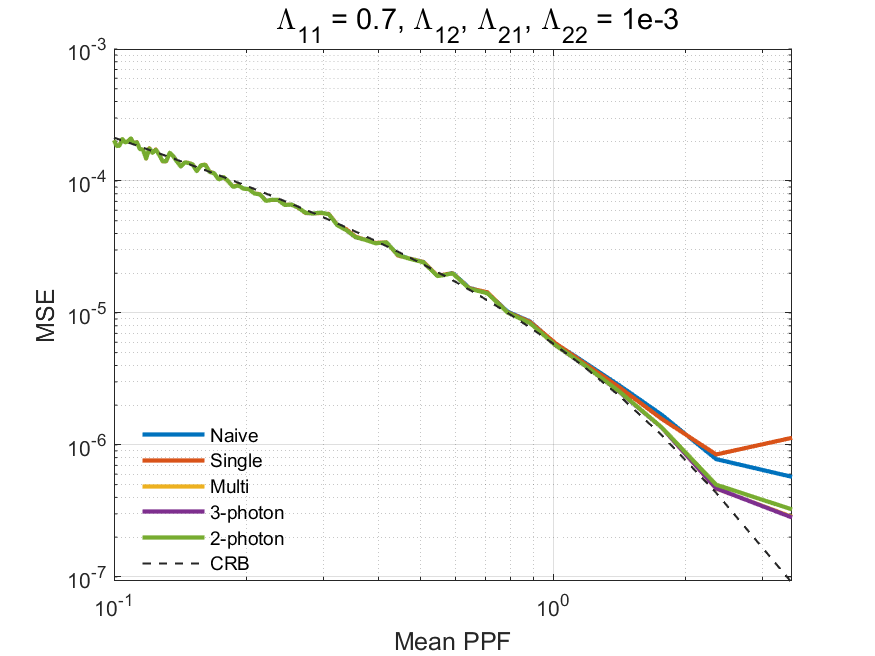} & 
        \includegraphics[width=0.33\textwidth]{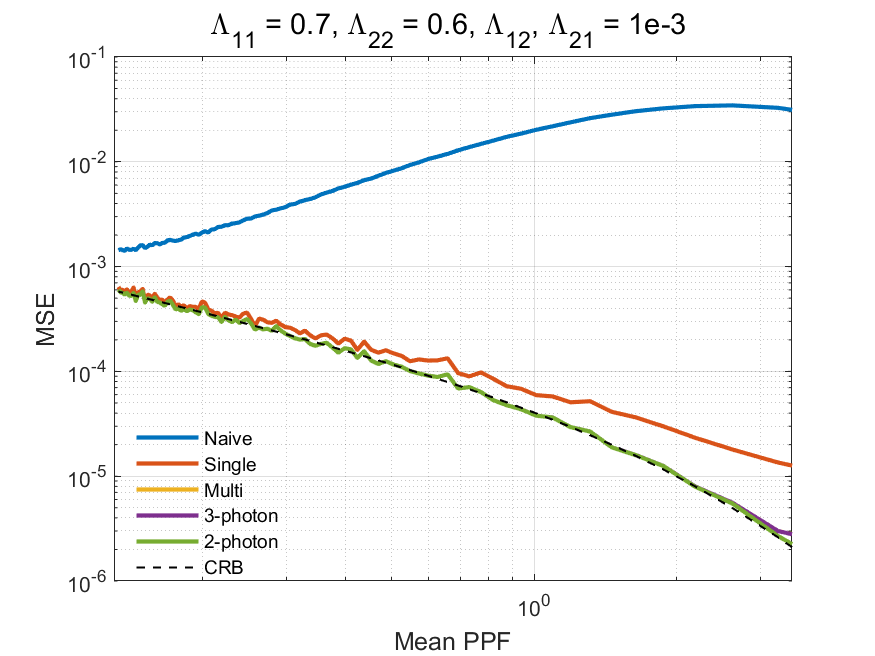} &
        \includegraphics[width=0.33\textwidth]{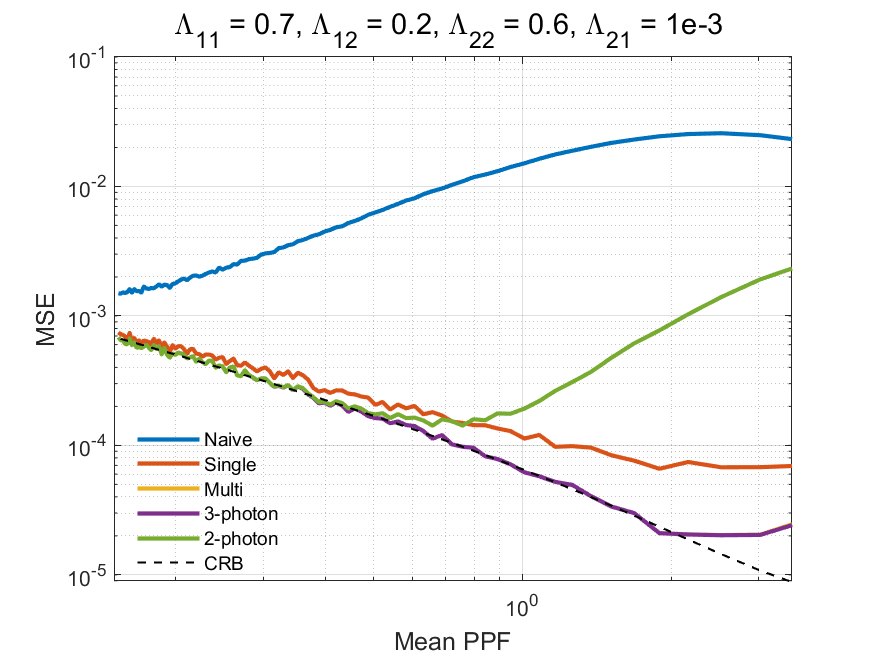}
    \end{tabular}
    \caption{ \centering Dependence of estimator performance on the ground truth image. The 4-photon estimator closely matches the 3-photon estimator in these cases.}
    \label{fig:CRB_Suppl}
\end{figure*}

The CRB curve in \cref{fig:CRB} is obtained using the mean of the diagonal elements of $\cI^{-1}$.
This curve and the MSEs of the estimators depend on the chosen ground truth $\Lambda$.
We illustrate this with three additional examples beyond the case shown in \cref{fig:CRB}.
When the ground truth has high flux at only two pixels, the 2-photon estimator matches the CRB closely across the range of PPF values studied as seen in \cref{fig:CRB_Suppl} (middle). This ground truth would mean that most ambiguous readouts arise from two-photon events. Thus, a 2-photon estimator reconstruction should closely match the ground truth. However, when the ground truth has high flux at three pixels as in \cref{fig:CRB_Suppl} (right), this model misattributes photon counts to just two pixels (due to the rejection of three- and four-photon terms in \eqref{eq:g_9}) resulting in a biased reconstruction. 

\section{Additional Comparisons}
\label{sec:Baselines}
\begin{figure*}

        \centering
        \includegraphics[width=0.9\textwidth, trim = {3cm, 2cm, 2.3cm, 2cm}, clip]{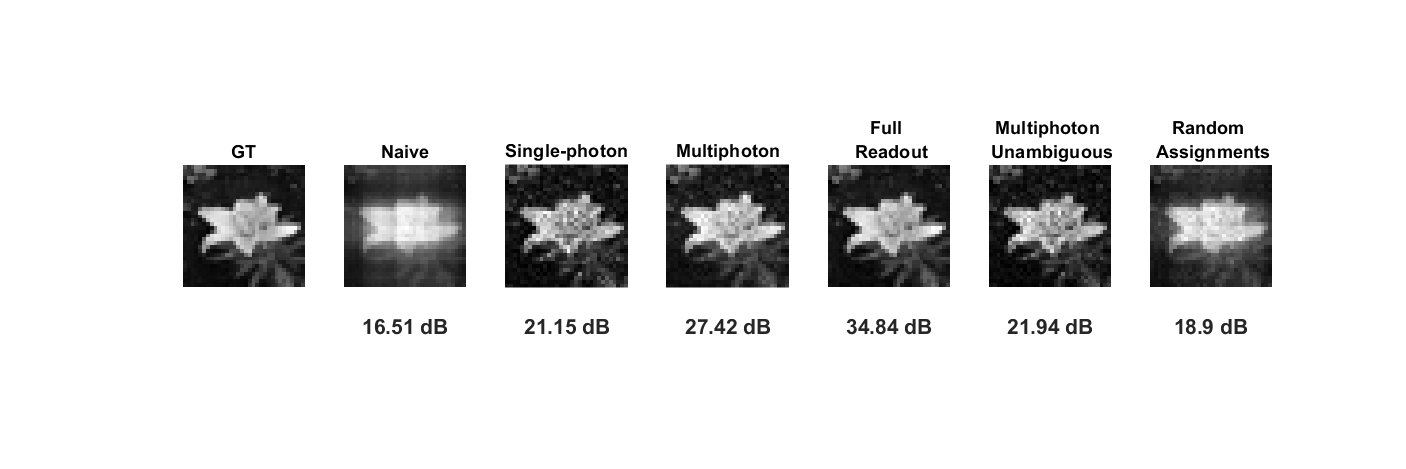} \\[5pt]
        \includegraphics[width=0.9\textwidth, trim = {3cm, 2cm, 2.3cm, 2cm}, clip]{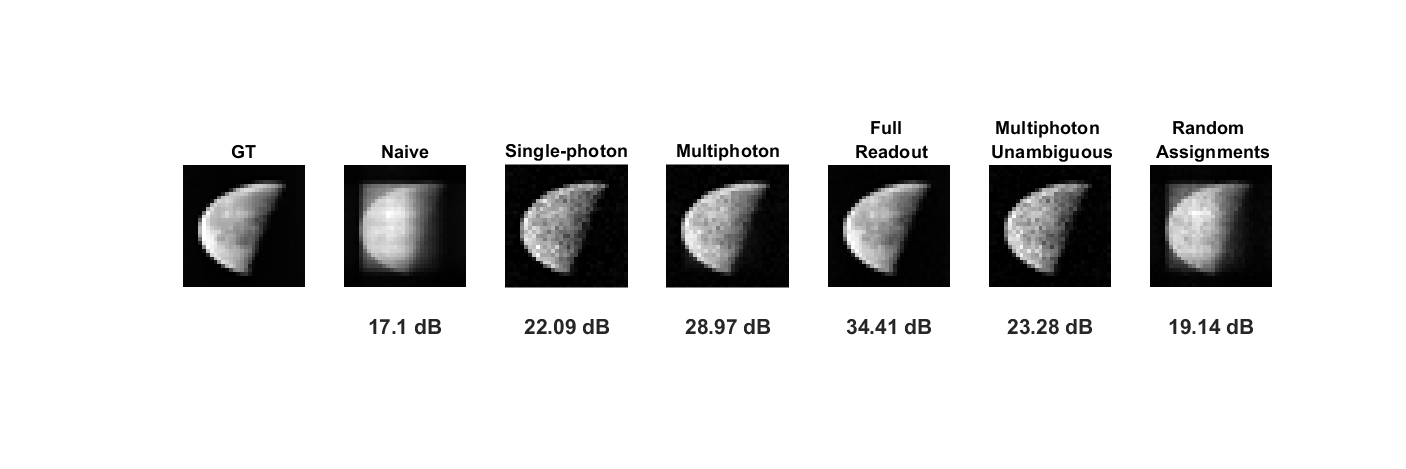} \\[5pt]
         \includegraphics[width=0.9\textwidth, trim = {3cm, 2cm, 2.3cm, 2cm}, clip]{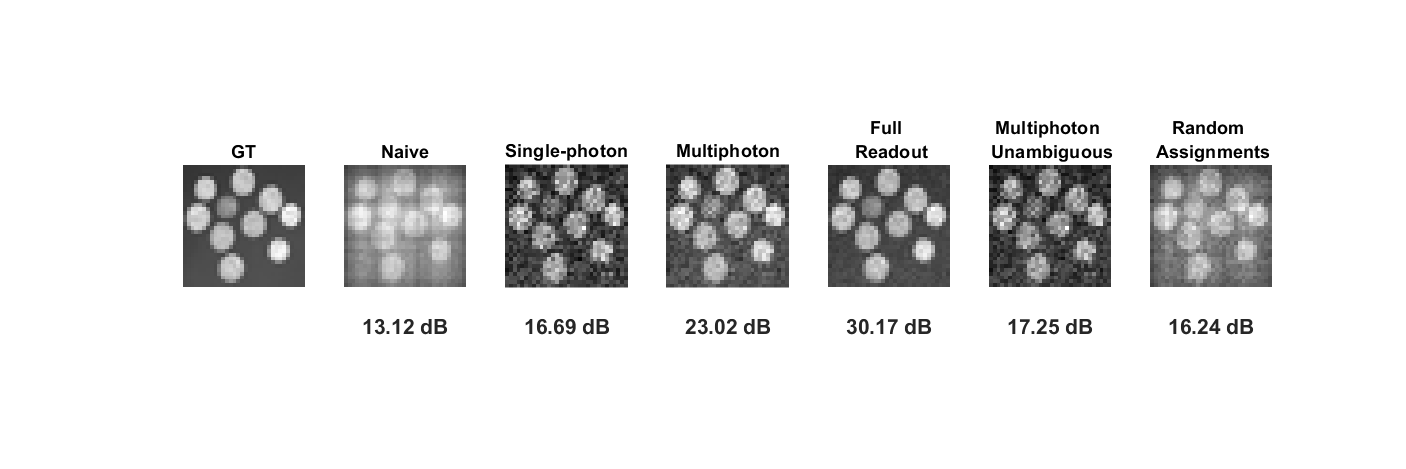}
   
        \caption{ \centering Comparison of the NE, SPE, and ME with full readout, multiphoton unambiguous estimator, and random assignments method.}
    \label{fig:baselines}
\end{figure*}

Here, we compare the performance of the ME with three additional baselines. The randomized assignments method is a modification of the NE where instead of imputing photon counts to \emph{every} candidate pixel where photon incidence could have occurred, the estimator randomly picks one solution from the set of possible photon incidence locations.
This leads to a decrease in the bias of the NE, as seen in the increase in PSNR in \cref{fig:baselines}.

Multiphoton events occurring along the same row or same column are unambiguous as noted in \cref{sec:img_model}. We can define a multiphoton unambiguous estimator that improves upon the SPE by discarding only ambiguous multiphoton events. Since this estimator uses more of the measured data and is unbiased, we expect its variance to be lower than the SPE\@. This is reflected in the increased PSNR value of reconstructions shown in \cref{fig:baselines}. 

Finally, we provide comparisons against a full readout which is free from ambiguities. We expect this model to only contain Poisson noise and hence have the best reconstruction among the methods we compare. 

It can be seen that across the baselines considered, our multiphoton estimator achieves the highest reconstruction PSNR\@.

\end{document}